\documentclass[twocolumn]{aastex63}
\usepackage[utf8x]{inputenc}
\usepackage{graphicx}
\usepackage{savesym} 
\savesymbol{tablenum} 
\usepackage{siunitx} 
\restoresymbol{SIX}{tablenum} 
\DeclareGraphicsExtensions{.pdf,.png}

\usepackage{natbib}
\received{October 12, 2020}
\revised{*** *, 2020}
\accepted{*** *, 2020}
\submitjournal{ApJ }

\shorttitle{Jet-driven piston shock}
\shortauthors{Maguire et al.}
\graphicspath{{./}{figures/}}

\begin{document}

\title{LOFAR observations of a jet-driven piston shock in the low solar corona}
   

\author{Ciara A. Maguire}
\affiliation{School of Physics, Trinity College Dublin, Dublin 2, Ireland.}
\affiliation{Astronomy \& Astrophysics Section, Dublin Institute for Advanced Studies, Dublin, D02 XF86, Ireland.}
\author{Eoin P. Carley}
\affiliation{Astronomy \& Astrophysics Section, Dublin Institute for Advanced Studies, Dublin, D02 XF86, Ireland.}

\author{Pietro Zucca}
\affiliation{ASTRON Netherlands Institute for Radio Astronomy, Dwingeloo, The Netherlands.}

\author{Nicole Vilmer}
\affiliation{LESIA, Observatoire de Paris, PSL Research University, CNRS, Sorbonne Universités, UPMC Univ. Paris 06, Univ. Paris Diderot, Sorbonne Paris Cité, 5 place Jules Janssen, 92195 Meudon, France.}
\affiliation{Station de Radioastronomie de Nançay, Observatoire de Paris, PSL Research University, CNRS, Univ. Orléans, 18330 Nançay, France.}

\author{Peter T. Gallagher}
\affiliation{Astronomy \& Astrophysics Section, Dublin Institute for Advanced Studies, Dublin, D02 XF86, Ireland.}
However, the precise origin of the associated shocks in the low corona is still subject to investigation

\begin{abstract}
The Sun produces highly dynamic and eruptive events that can drive shocks through the corona. These shocks can accelerate electrons, which result in plasma emission in the form of a type II radio burst. Despite the large number of type II radio bursts observations, the precise origin of coronal shocks is still subject to investigation. Here we present a well observed solar eruptive event that occurred on 16 October 2015, focusing on a jet observed in the extreme ultraviolet (EUV) by the Atmospheric Imaging Assembly (SDO/AIA), a streamer observed in white-light by the Large Angle and Spectrometric Coronagraph (SOHO/LASCO), and a metric type II radio burst observed by the LOw Frequency Array (LOFAR). LOFAR interferometrically imaged the fundamental and harmonic sources of a type II radio burst and revealed that the sources did not appear to be co-spatial, as would be expected from the plasma emission mechanism. We correct for the separation between the fundamental and harmonic using a model which accounts for scattering of radio waves by electron density fluctuations in a turbulent plasma. This allows us to show the type II radio sources were located $\sim$0.5 R$_\odot$ above the jet and propagated at a speed of $\sim$1000\,km\,s$^{-1}$, which was significantly faster than the jet speed of $\sim$200\,km\,s$^{-1}$. This suggests that the type II burst was generated by a piston shock driven by the jet in the low corona.

\end{abstract}


\section{Introduction} \label{sec:intro}
The Sun regularly produces a variety of highly dynamic and energetic explosive events such as coronal  mass  ejections (CMEs), flares, erupting loops or plasmoids, ejecta-like sprays and jets \citep{Klein1999,Daupin2006,Zimovets2012,Carley2013,Morosan2018,Maguire2020, Chrysaphi2020}. The mass motions during these eruptive events can often travel with speeds that exceed the local background Alfvén speed, which result in the formation of plasma shocks. The acceleration of electrons at the shock front can prompt coherent plasma emission at both the fundamental ($f_p$) and second harmonic ($2f_p$) of the plasma frequency \citep{Nelson1985a,VrsnakCliver2008}. The radio emission produced in this process is referred to as a type II radio burst and since the plasma frequency $f_p$ is dependent upon the background electron density $n_e$ via $f_p \approx 9000\sqrt{n_e(cm^{-3})}$ MHz, type II bursts provide a useful diagnostic of local coronal conditions and shock parameters. Furthermore, observations of type II bursts can provide insight into the origin of coronal shocks and help us determine whether they are (1) flare related due to blast waves or (2) CME or small scale ejecta related \citep{Zimovets2012, Eselevich2015, Mancuso2019}. The shock can be further classified as a bow shock or a piston-driven shock. For the bow shock scenario, the ambient plasma is able to flow around the driver so that the shock and driver are seen to propagate at the same velocity \citep{Cho2007,Schmidt2016}. While in the case of a piston-driven shock, the plasma is unable to flow behind the driver so that the distance between the driver and shock increases with time and the shock speed can be several times that of the driver \citep{Pomoell2008,Nindos2011,Bain2012,Grechnev2018}. 

To date, the origin of plasma shocks have predominantly been studied in terms of highly energetic events namely; strong flares \citep{Zucca2018b}, X-ray jets \citep{Klein1999}, erupting coronal loops \citep{Daupin2006}, eruptive magnetic flux rope \citep{Wang2017},  plasmoids \citep{Bain2012,Carley2013} and CMEs \citep{Maguire2020}. However few studies have investigated type II bursts associated with EUV jets and weak CMEs (see \cite{Chrysaphi2020} as an example). Here we present observations of a C-class flare and a narrow jet that resulted in a metric type II radio burst. We determine the location of the type II burst and carry out a multi-wavelength kinematic analysis to infer the origin of the shock.

\begin{figure}[t!]
\includegraphics[scale=0.43, trim=1cm 1cm 1cm 0cm]{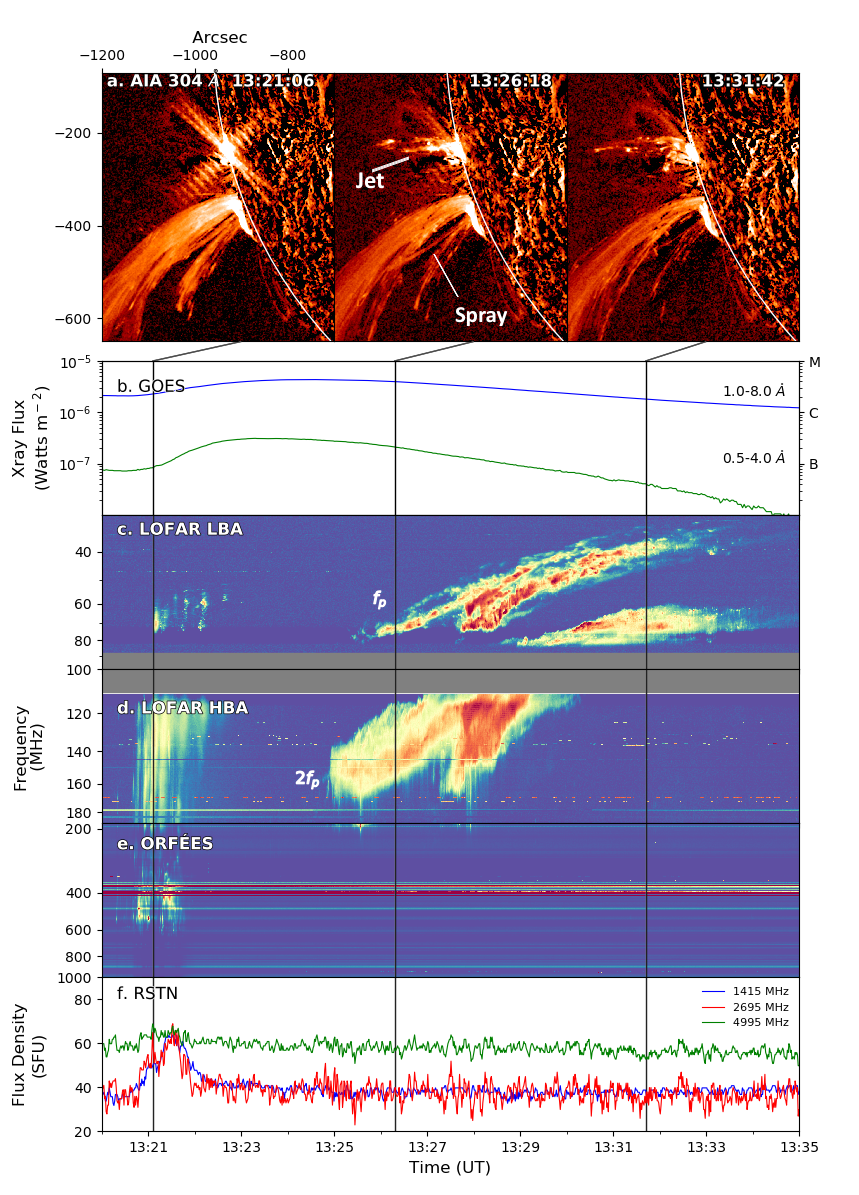}
\caption{(a) Base  difference  images  of  the  jet and spray observed  with  AIA  304  \r{A} ,  (b)  GOES  0.5-4  \r{A}  and  1-8 \r{A} soft X-ray flux of the C4.3 class solar flare. The remaining panels show the radio emission as observed by (c) LOFAR’s low band antennae (LBA) (30-90 MHz), (d) LOFAR’s high band antennae (HBA) (110-240 MHz), (e) ORFÉES (140-1000 MHz) and (f) RSTN channels ( 1415, 2695, 4995 MHz). The LOFAR dynamic spectra shows a type II radio burst with fundamental ($f_p$) and harmonic ($2f_p$) components initiating at 13:25 UT and ceasing at 13:34 UT.
}
\label{fig1}
\end{figure}

\begin{figure*}[t!]
\includegraphics[scale=0.51, trim=1cm 1cm 2cm 2cm]{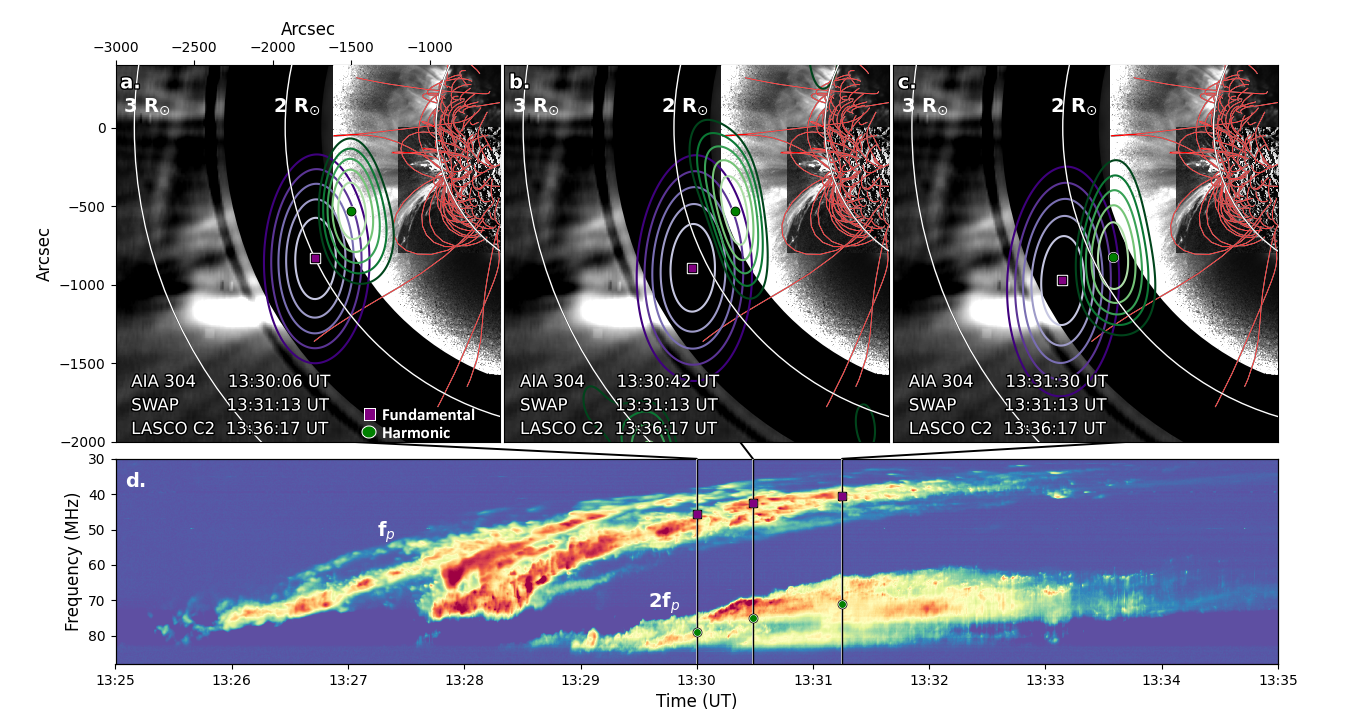}
\caption{ (a-c) Type II radio burst observed by LOFAR at three separate times. The purple and green contours represent 50-90$\%$ the peak flux density of the fundamental ($f_p$) and harmonic ($2f_p$) radio sources, respectively. The purple square and green dot represent the bursts' centroid position. The burst contours are overlaid on composite images from AIA 304 \r{A} images (innermost), SWAP 174 \r{A} (central) and LASCO C2 (outermost).  The coronal magnetic field determined from the PFSS is shown by red lines. The solid white circles indicate distances of 2 and 3 R$_\odot$. (d) The corresponding dynamic spectrum showing the $f_p$ and $2f_p$ components. Purple squares and green dots denote the points along the burst that have been imaged. 
}
\label{fig2}
\end{figure*}

Our kinematic analysis includes an investigation of low frequency radio wave scattering in the corona, which is necessary to account for radio source displacements from their true position. Early observations by the Culgoora Radioheliograph revealed that type II fundamental emission is radially shifted outwards with respect to harmonic emission \citep{ Kai1968,Sheridan1972,NelsonSheridan1974,NelsonRob1975,Suzuki1985}. Such behaviour is attributed to radio wave scattering \citep{  Fokker1963,Steinberg1971,Stewart1972,Riddle1974,Bastian1994}. More recently, LOw Frequency Array \citep[LOFAR;][]{vanHaarlem2013} tied-array beam observations demonstrated that band-split type II fundamental sources experience displacement due to radio wave scattering in a turbulent medium \citep{Chrysaphi2018}. In this study, we use LOFAR interferometric observations, which have superior spatial resolution with respect to tied-array observations, to image the separation between type II fundamental and harmonic components with unprecedented spatial and temporal resolution. We account for the spatial displacement between fundamental and harmonic sources using a model of radio wave scattering in the corona which allows for a necessary correction of radio source positions and their comparison with the shock driver imaged in EUV, showing they follow the kinematics of a piston-driven shock. 

In Section \ref{sec:obs}, observations of the flare, jet and type II radio burst are presented. The observational method and models used to determine the radio source location are described in Section \ref{Analysis}. We discuss the shock origin and the nature of radio wave scattering in Section \ref{Discussion}, and finally, conclusions are presented in Section \ref{conclusion}.

\section{Observations} 
\label{sec:obs}

A GOES C4.3 class flare (Figure \ref{fig1}b) began 2015 October 16 at ∼13:20 UT from active region NOAA 12435 (SOL2015-10-16T13:25:30). The flare was located on the solar eastern limb and inspection of data from the Extreme Ultraviolet Imager \citep[EUVI;][]{Wuelser2004} onboard the STEREO-A spacecraft near the time of the flare revealed that the active region extended around the far side of the Sun. Base difference images shown in Figure \ref{fig1}(a) from the 304 \r{A} passband of the Atmospheric Imaging Assembly \citep[AIA;][]{Lemen2012} onboard the Solar Dynamics Observatory \citep[SDO;][]{Pesnell2015} illustrate the evolution of a jet that emerged from the active region during the impulsive phase of the flare. The jet originates from a foot point on the limb, meaning it most likely propagated close to the plane-of-sky (POS). The ejected material initially moved radially before moving slightly southward. Below the jet, a spray-like feature was observed to propagate in a southward direction. The spray emerged an hour prior to the flare and persisted for the duration of the eruption.

In Figure 1 (c) to (f) the spectral radio observations from various ground instruments are shown, namely LOFAR's remote station RS509 observing between 10-240 MHz, the radio spectrograph Observation Radio Frequence pour l’Étude des Eruptions Solaires (ORFÉES) \footnote{http://secchirh.obspm.fr/spip.php?article19}, observing between 140-1000 MHz and the Learmonth site of the Radio Solar Telescope Network (RSTN)\footnote{http://www.ngdc.noaa.gov} measuring solar radio flux density. Coinciding with the onset of the GOES X-ray at 13:20 UT, a group of type IIIs were detected by LOFAR and ORFÉES, as shown in panels (c) to (e). Subsequently at $\sim$13:25 UT, LOFAR observed a strong type II radio burst with well defined fundamental and first harmonic emission bands, indicated in Figure \ref{fig1}(c \& d) by $f_p$ and $2f_p$, respectively. Both the fundamental and harmonic emission bands exhibit detailed structure, band splitting and fragmentation into multiple bands with different drift rates. At the time of the type II burst there was no significant radio emission above 200 MHz (see panel e $\&$ f), which suggests that no radio emission was generated or escaped from low in the corona and that the flare may have been partially occulted. Unfortunately there were no STEREO A images at the time of the eruption to observe the evolving active region.

LOFAR also provided interferometric observations of the event until 14:00 UT using the low band antennas (10-90\,MHz) from 36 stations (24 core and 16 remote). The maximum baseline of the LOFAR observation was 84\,km, which gave sub-arcminute resolution across almost all of the observed frequency range. Observations of the calibrator source, Virgo A, were taken simultaneously over all subbands. The visibility data was recorded with a correlator integration time of 0.167 s \citep{Zhang2020}. The data was processed using the Default Processing Pipeline \citep[DPPP;][]{vanDiepen2018} followed by an implementation of WSCLEAN \footnote{https://gitlab.com/aroffringa/wsclean/} to produce images with a spectral resolution of 195.3 kHz and cadence of 1 second. 

\section{ Data Analysis and Results}
\label{Analysis}
In the following, we determine the location of the fundamental and harmonic components of the type II burst in relation to the jet observed in EUV. This can only be done after accounting for radio wave propagation effects, allowing us to determine where the radio burst was generated in relation
to the eruptive structure and the kind of coronal environment that lead to shock formation.

\subsection{Imaging of radio burst}
In order to track the motion of the shock, we image the fundamental and harmonic components of the type II burst at multiple moments in time. Figure \ref{fig2}(a-c) illustrates the position of the fundamental (purple contours) and harmonic (green contours) component of the type II burst overlaid on composite images from AIA 304 \r{A} (innermost), Sun Watcher using Active Pixel  \citep[SWAP;][]{Berghmans2006} 174 \r{A} (central) and LASCO C2 (outermost). The red lines represent the Sun’s coronal magnetic field, which was extrapolated from the photospheric magnetic field using the Potential-Field Source-Surface  model \citep[PFSS;][]{Stansby2019} with data from the Global Oscillation Network Group \citep[GONG;][]{Harvey1996}. Figure \ref{fig2}(d) demonstrates the type II radio burst dynamic spectrum with the fundamental and harmonic emission bands labelled as $f_p$ and $2f_p$, respectively. The purple squares and green dots indicate the points imaged along the burst as seen in panels a-c. The imaging reveals that the fundamental source (purple contours) is radially shifted outwards with respect to the harmonic source (green contours) by 0.3-0.5 $R_{\odot}$. We find that regardless of where we image on the emission bands at one particular time, there is a clear separation between the fundamental and harmonic sources. Such behaviour contradicts the underlying plasma emission mechanism according to which fundamental and harmonic radio waves are generated in the same location and should therefore appear co-spatial \citep{Melrose1975}. The observed displacement is potentially due to the scattering of radio waves by electron density fluctuations that exist due to turbulent plasma processes in the corona \citep{Steinberg1971,Stewart1972, Riddle1974, NelsonSheridan1974}. Scattering effects are particularly significant on fundamental (as opposed to harmonic) radio waves because the fundamental emission is close to the plasma frequency and therefore strongly affected by propagation effects, e.g., due to small-scale variations in the background density of the plasma. This variation in the background density determines the level of scattering of radio waves and is described by the relative level of root mean squared (r.m.s) density fluctuations $\varepsilon = \sqrt{\left< \delta n^2 \right>}/n$, where $n$ is the electron density. 

 In the next section we account for the effects of scattering on fundamental emission to correctly interpret the type II observations for this event and in the process gain insight into the parameters that describe radio wave scattering.

\begin{figure*}[t!]
\begin{center}
   \includegraphics[scale=.68, trim=1.5cm 1cm 2cm 1.5cm]{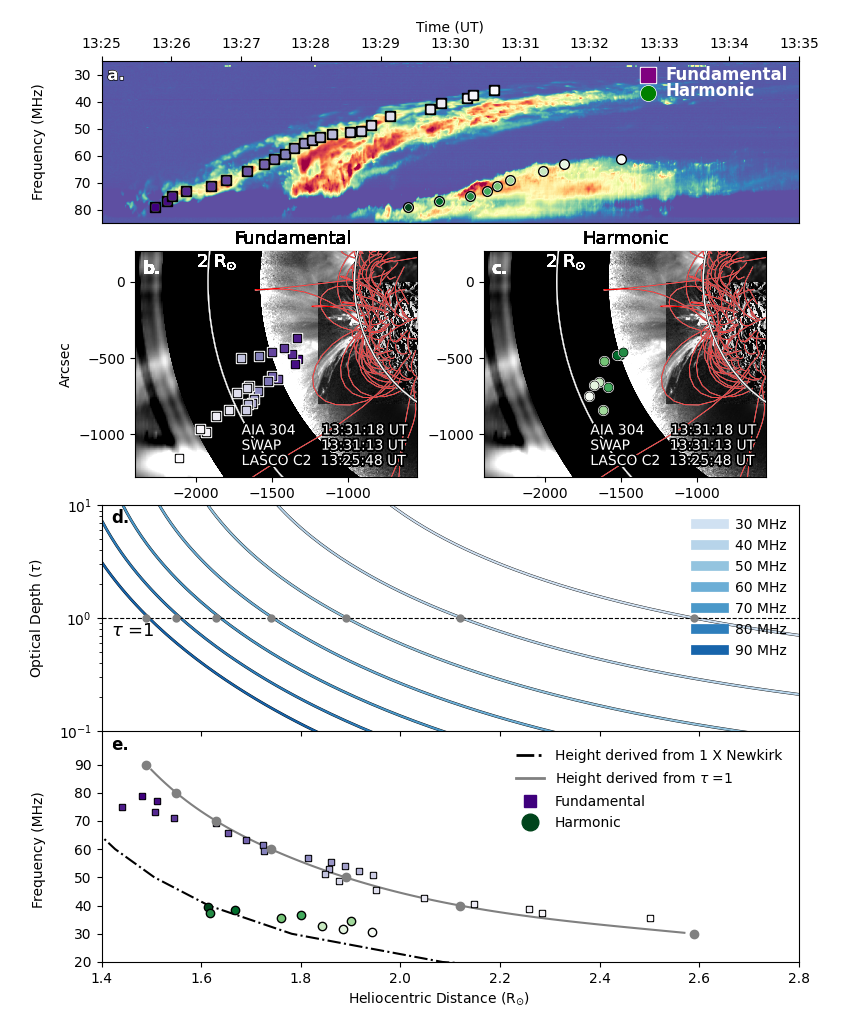} 
\end{center}
\caption{(a) LOFAR dynamic spectrum depicting the fundamental ($f_p$) and harmonic emission ($2f_p$) of a type II burst. The purple squares and green circles overlaid illustrate the frequency and times of the burst that were imaged, with dark-to-light shading representing progression in time. (b $\&$ c) show the centroids of the fundamental (purple squares in b) and harmonic component (green circles in c) of the type II burst overlaid on an EUV and WL composite image from SDO/AIA 304 \r{A}, PROBA2/SWAP 174 \r{A} and SOHO/LASCO C2. The coronal magnetic field determined from the PFSS is shown by red lines. (d) The optical depth with respect to scattering for radio waves as a function of heliocentric distance for a range frequencies between 30 and 90 MHz. The dashed line indicates $\tau$ =1 and the grey dots indicate the height at which the radio source is predicted to appear for each frequency. (e) The heights of fundamental (purple squares) and harmonic (green dots) sources as observed by LOFAR.The grey line represents where the scattering model predicts the fundamental radiation eventual escapes and the dashed black line represents where the emission is generated, as predicted by the Newkirk model.
}
\label{fig3}
\end{figure*}

\subsection{Scattering of fundamental plasma emission}
\label{scatter_model}
The dynamic spectrum presented in Figure \ref{fig3}(a) shows the type II fundamental and harmonic emission bands marked with purple squares and green circles, respectively, with dark-to-light shading representing progression in time. The time-frequency points were selected using a fundamental to harmonic frequency ratio of 1.8-1.9, to be consistent with observations \citep{Melink2018}. Panels (b) and (c) depict the positions of the fundamental and harmonic sources on a composite image from AIA 304 \r{A}, SWAP 174 \r{A} and LASCO C2. The coronal magnetic field determined from the PFSS is shown by red lines. Here we assume the displacement between the fundamental and harmonic emission is caused by radio wave scattering. To estimate the extent to which the fundamental is shifted by scattering, we adopt the \cite{Chrysaphi2018} model. This model assumes that as radio waves propagate through the corona they undergo repeated small-angle deflections due to isotropic fluctuations in the plasma density caused by turbulence (see \cite{Gordovskyy2019} for details). The optical depth with respect to scattering for radio waves in the corona is described as 
\begin{equation}
    \tau(r) = \int_{r}^{1 AU} \pi\frac{f_{p}^4(r)}{(f^2 -f_{p}^2(r))}\frac{\varepsilon ^ 2}{h}  dr
\label{tau}
\end{equation}
where $f_p$ is the plasma frequency, \textit{h} is the effective scale length of density fluctuations and $\varepsilon$ is relative level of electron density fluctuations\footnote{Eq.\ref{tau} is adopted from Eq.9 of \cite{Chrysaphi2018}, where we have assumed power-law spectrum of electron density fluctuations, which is more consistent with in-situ observations \citep{Bastian1994}. This means that the coefficient $\sqrt(\pi)/2$ is now $\pi$ (following Eq.31 of \cite{Thejappa2007} and Eq.34 of \cite{Thejappa2008}}. A given model of $f_p$ predicts where the emission is generated, and where we expect to see harmonic emission, since it undergoes very little scattering. The Newkirk model best describes the positions of the harmonic sources, assuming the shock propagated close to POS (see Figure \ref{fig3}e). Considering Equation \ref{tau}, $\tau(r) = 1 $ corresponds to the heliocentric distance at which fundamental radio emission is expected to escape. The value of $\varepsilon^2/h$ was obtained from optimising the fit between the heights predicted at $\tau$=1 with the radial positions of the fundamental emission. Using this approach $\varepsilon^2/h$ was found to be 2$\times$10$^5$\,km$^{-1}$.

Figure \ref{fig3}(d) illustrates the solution to Equation \ref{tau}, showing how $\tau$ varies with $r$ for different values of $f$ ($f$ in range 30-90 MHz in steps of 10 MHz). The dashed line indicates the point at which $\tau(r)$ = 1. The expected height of scattered fundamental emission at each frequency is marked by a grey dot. In Figure \ref{fig3}(e), the grey line represents where the model predicts the fundamental radiation eventual escapes. The dashed black line is where the emission is generated (according to the Newkirk model), and where the harmonic emission should be observed. The heights of fundamental sources (purple squares) agree quite well with the scattering model (grey line) while the harmonic sources (green dots) are in agreement with the Newkirk model (dashed black line). This shows the spatial displacement of these radio sources is accounted for by the scattering model. 

It should be noted that there is a deviation between the models and data at higher frequencies and this may be an effect of the Newkirk model's inability to accurately describe the complex structure of the low corona. There are a plethora of density models such as Mann, Baumbach–Allen and Saito, however these models predict even lower densities at these heights \citep{Mann1999b,Baumbach1937,Allen1947, Saito1970}. The Newkirk model was therefore established as the most appropriate to describe the observed source positions. 

To summarize, LOFAR provided images of the fundamental and harmonic emission so that we were able to identify where the radio waves were generated (location of harmonic) as well as where the scattered radio waves escaped (location of fundamental). Overall the \cite{Chrysaphi2018} model successfully accounts for the spatial separation between the fundamental and harmonic emission. The model proved to be a reliable means to correct for the positional shift due to scattering, so that we can accurately determine the type II burst kinematics.

\subsection{Kinematics of ejecta and type II radio burst}

\begin{figure}[t!]
\includegraphics[scale=.6, trim=  .5cm .5cm .5cm  .5cm]{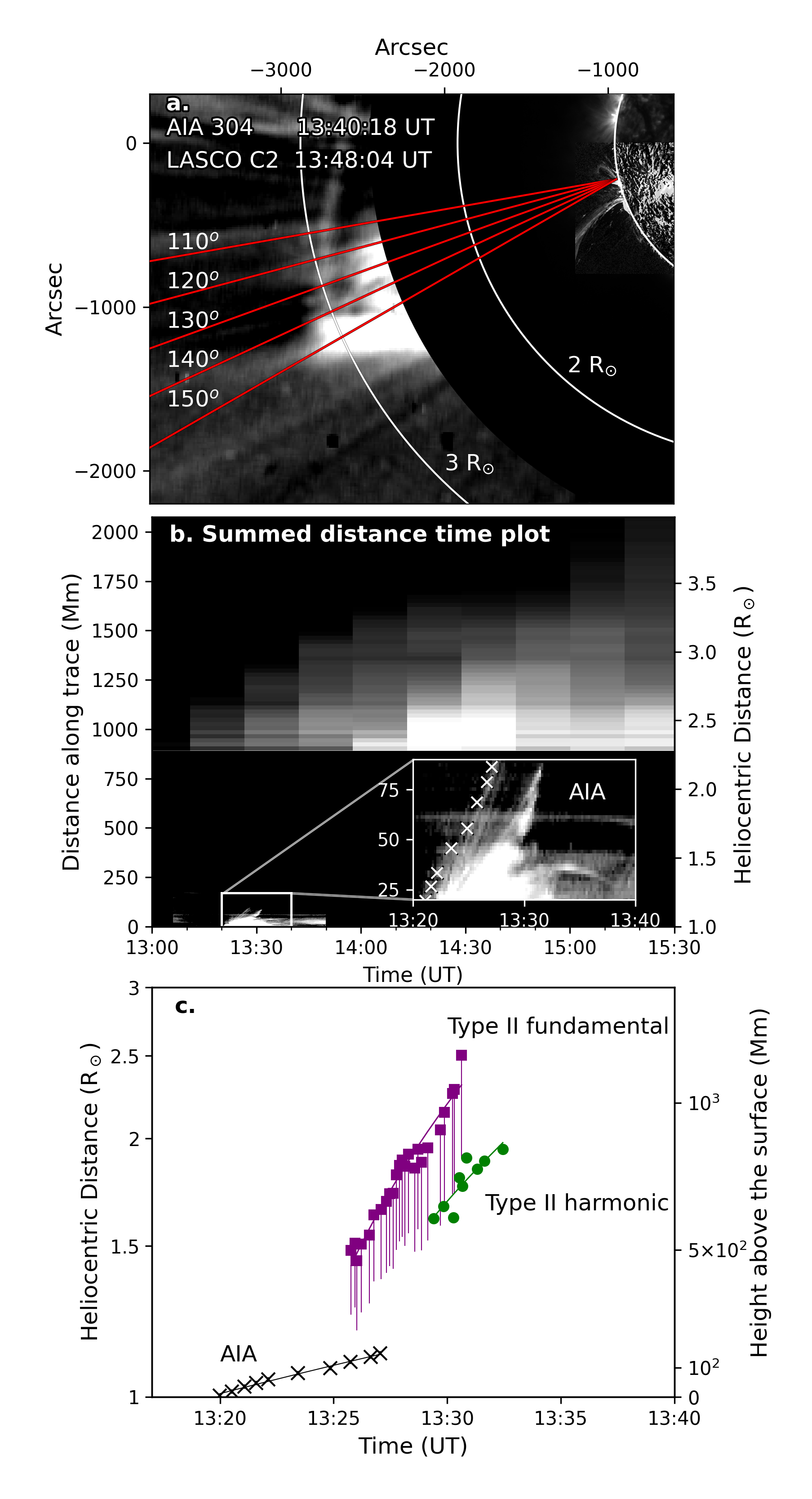}  
\caption{(a) Composite of base difference images from AIA 304 \r{A} and LASCO C2, with five red traces over the region of interest. The traces start \ang{110} to the solar north and are separated by \ang{10}. (b) The sum of the time distance plots generated along the five traces and a zoom in of the AIA field of view.  (c) The height time profiles of the EUV and radio features. The jet as seen by AIA is marked by black crosses and the position of the fundamental and harmonic type II radio emission are marked by purple squares and green dots, respectively. The error bars associated with the fundamental emission represent the scattering induced radial displacement. An animation of the LASCO C2 base difference images is available in the online Journal. The animation runs from 12:12 to 17:48 UT.}
\label{fig4}
\end{figure}
Figure \ref{fig4}(a) presents a composite of base difference images from SDO/AIA 304 \r{A} and LASCO C2 showing the jet on the solar limb and an over arching helmet streamer situated $\sim$\ang{100} to solar north. To determine the ejecta kinematics, five traces were examined around the region of interest, indicated by five red lines in Figure \ref{fig4}(a). The traces originate at the active region from the solar limb, starting \ang{110} to the solar north and are separated by \ang{10}. The distance-time plot associated with each of the traces was summed to produce the plot illustrated in Figure \ref{fig4}(b). The inset is a zoom in on the AIA FOV showing the jet, which appears to have a few components, as indicated by the two prong structure. We take the foremost component as a measure of the jet front. The crosses overlaid on Figure \ref{fig4}(b) indicate the points selected using a point-and-click technique. The fundamental and harmonic emission heights were taken as the distance between source centroid to solar center. Combing the EUV and radio data, a height time profile was constructed as shown in Figure \ref{fig4}(c). The jet is marked by black crosses and the height of the type II  fundamental and harmonic emission are marked by purple squares and green dots, respectively. The error bar associated with the fundamental emission represents the scattering induced radial shift, as calculated in Section \ref{scatter_model}. The error associated with the EUV heights was deduced from 10 trial measurements of height in Figure\ref{fig4}(b) and was found to be $\sim$0.1 $R_{\odot}$ ($\sim$10 pixels). The jet was observed to have an average velocity of $\sim$200\,km\,s$^{-1}$ and the type II fundamental and harmonic travelled at $\sim$1000\,km\,s$^{-1}$ and $\sim$1090\,km\,s$^{-1}$, respectively. The significance of these results are discussed in Section \ref{shock_origin}. We note that although there is slight movement in the streamer observed by LASCO C2 (see online animation associated with Figure \ref{fig4}), it is unclear whether this is associated with the motion that lead to the type II burst therefore we concentrate on the kinematics of the jet and type II burst in this study. 
\section{Discussion} \label{Discussion}

\subsection{What is the origin of the shock?}
\label{shock_origin}
As seen in Figure \ref{fig1}, the EUV jet emerged at 13:20 UT from the solar limb and propagated outwards at a speed of $\sim$200\,km\,s$^{-1}$. Although the jet initially moved radially, the PFSS in Figure \ref{fig3}(b$\&$c) suggests that the ejected material later moved southward due to the closed magnetic field lines. About five minutes later, the type II burst was observed $\sim0.5R_{\odot}$ above the jet and had a significantly larger velocity ($\sim$1000\,km\,s$^{-1}$), which is indicative of a piston-driven shock 
\citep{Maxwell1985,Liepmann1957,Pomoell2008, VrsnakCliver2008}.  We note that we have based this analysis on the assumption that the driver and shock propagate close to the plane-of-sky, even if this was not precisely the case our interpretation is still valid. For example, if the shock propagated at an angle of \ang{20} from the plane of sky, the shock speed would still exceed the speed of the driver, which is characteristic of a piston-driven shock.

We found the local Alfvén speed in region of interest by combining the Newkirk electron density model and a 2D plane-of-sky magnetic field map (derived from a PFSS model). Considering the same five traces in Figure \ref{fig4}(a), the average Alfvén speed at 2 $R_{\odot}$ was found to be 740$\pm$70\,km\,s$^{-1}$. The fact the jet (shock-driver) propagated at sub-Alfvénic velocities ($\sim$200\,km\,s$^{-1}$), provides further evidence that the shock was piston-driven \citep{VrsnakCliver2008}. The Alfvén Mach number $M_A$ of the shock was estimated to be $\sim$1.35 by taking the ratio of the shock speed ($\sim$1000\,km\,s$^{-1}$) to Alfvén speed (740$\pm$70\,km\,s$^{-1}$). This $M_A$ value is  consistent with previous studies \citep{Vrsnak2001,Zucca2014,Maguire2020}. The $M_A$  was also estimated from the band splitting seen in the type II fundamental emission band at 13:27:30 - 13:30:00 UT. Using the relative instantaneous bandwidth between the upper and lower split bands, the compression ratio  $X$  was found to be in the range 1.3-1.5. To determine $M_A$ values, we used the expression from \citet{Vrsnak_Mag2002} for a perpendicular shock:
\begin{equation}\label{eq:Ma}
M_A = \sqrt{\frac{X(X + 5 +5\beta)}{2(4-X)}}
\end{equation}
where $\beta$ is the plasma-to-magnetic pressure ratio ($\beta$ $<<$1). The values for $M_A$ were found to be 1.3-1.4, which is consistent with $M_A$  derived from the shock speed to Alfvén speed ratio.

In summary we suggest that as the jet erupted, a piston-driven shock was established ahead of it and the streamer may have acted as a tube for the shock to propagate down \citep{Eselevich2015}. Piston-driven shocks with type II emission have often been associated with wide and fast CME drivers \citep{Kahler2019} but few have reported piston shocks resulting from narrow ejecta low in the corona as is the case in this event.

\subsection{ Radio wave scattering in the low corona}
\label{scatter_discuss}
In Section \ref{scatter_model} we showed that the scattering model successfully accounts for the spatial separation between the type II fundamental and harmonic emission. Let us consider the validity of the model's assumptions, namely, that scattering is the dominant radio wave propagation effect in the low corona and that scattering is due to isotropic density fluctuations.

We provide evidence that scattering is the dominant propagation effect on radio waves by comparing the size of the fundamental and harmonic sources, normalised to the point spread function of LOFAR (see Figure \ref{fig2} a-c). The fundamental sources were found to be 1.6-1.9 times larger than the harmonic sources. This is as expected since scattering from density fluctuations is known to have a more significant effect on the fundamental emission rather than the harmonic \citep{NelsonRob1975, LengyelFrey1985}. 

The model used in this work also assumes radio wave scattering by isotropic density fluctuations. However, previous work suggests that density fluctuations are in fact anisotropic, which would imply $\varepsilon^2/h$ has both a parallel and perpendicular component \citep{Armstrong1990,Anantharamaiah1994}. In order to determine whether this assumption changes our results, we consider the effects of anisotropic scattering on radio sources. Numerical models by \cite{Kontar2019} suggest the radial shift experienced by a radio source due to anisotropic scattering is slightly less than in the isotropic scenario ($\sim$0.52 R$_\odot$ compared to $\sim$0.68 R$_\odot$ for a source propagating in POS). To account for the displacement under anisotropic scattering conditions, the values of $\varepsilon^2/h$ would have to be slightly larger. It is important to note that although anisotropy does not have a dramatic effect on the radial shift, it does affect source morphology, e.g. sources are expected to elongate perpendicular to the heliospheric radial direction due to enhanced scattering perpendicular to the large-scale (radial) magnetic field of the Sun or elongated sources \citep{Ingale2015,Kontar2017}.

While the simple analytical model used in this analysis can successfully account for shifted positions of the radio sources, it can not account for all observed properties of scattered sources (for example size and morphology). As such, future studies that combine fully developed numerical scattering models with interferometric observations from LOFAR are needed to comprehensively understand radio wave propagation in the turbulent plasma near coronal shocks.

\section{Conclusion}
\label{conclusion}
We present a study of a flare, jet and type II radio burst that occurred on 16 October 2015 on the eastern limb of the Sun. The purpose of this study was to determine the location of the type II burst and the origin of the associated plasma shock. We carried out a multi-wavelength kinematic analysis, which included an investigation of low frequency radio wave scattering in the corona. LOFAR interferometrically imaged both the fundamental and harmonic emission of a metric type II and revealed that the sources are not co-spatial, as would be expected from the plasma emission mechanism. We account for their spatial displacement using a model of radio scattering in the corona. This model allowed for necessary correction of source positions and their comparison with the shock driver. Furthermore, optimisation of the model to the data provided information about scattering parameters in particular the level of density fluctuations in the turbulent corona e.g. we found that $\varepsilon^2/h$  $\sim$2$\times$10$^5$\,km$^{-1}$, which is slightly lower compared to previous studies \citep{Chrysaphi2018}.  

After accounting for radio wave scattering effects, we determined where the radio burst was generated in relation to the eruptive structure and the coronal environment that lead to shock formation. We found that the type II burst was located at a much higher altitude than the EUV jet and had a significantly larger velocity, namely the jet speed was $\sim$200\,km\,s$^{-1}$ while the type II burst propagated at $\sim$1000\,km\,s$^{-1}$.The association of the sub-Alfvénic jet with the type II burst and the relative velocities of the jet and the type II emission provides strong evidence of a shock that was initially piston-driven.

\section*{Acknowledgments and Data Availability}
\indent C. A. M. is supported by an Irish Research Council Government of Ireland Postgraduate Scholarship. We are grateful to the GOES, AIA, SWAP and LASCO teams for open access to their data. The data sets generated during and/or analyzed during the current study are available in the LOFAR Long Term Archive, \url{https://lta.lofar.eu/} and \url{https://sdo.gsfc.nasa.gov/data/}. ORFÉES is part of the FEDOME project, partly funded by the French Ministry of Defense. We would like to thank Ambassade de France for providing a High level scientific mobility grant to allow for part of this work to be carried out in LESIA, Observatoire de Paris, France. 
\bibliographystyle{aasjournal}
\bibliography{sample63.bbl}

\begin{thebibliography}{}
\expandafter\ifx\csname natexlab\endcsname\relax\def\natexlab#1{#1}\fi
\providecommand{\url}[1]{\href{#1}{#1}}
\providecommand{\dodoi}[1]{doi:~\href{http://doi.org/#1}{\nolinkurl{#1}}}
\providecommand{\doeprint}[1]{\href{http://ascl.net/#1}{\nolinkurl{http://ascl.net/#1}}}
\providecommand{\doarXiv}[1]{\href{https://arxiv.org/abs/#1}{\nolinkurl{https://arxiv.org/abs/#1}}}

\bibitem[{Allen(1947)}]{Allen1947}
Allen, C.~W. 1947, Monthly Notices of the Royal Astronomical Society, 107,
  426–432

\bibitem[{Anantharamaiah {et~al.}(1994)Anantharamaiah, Gothoskar, \&
  Cornwell}]{Anantharamaiah1994}
Anantharamaiah, K.~R., Gothoskar, P., \& Cornwell, T.~J. 1994, Journal of
  Astrophysics and Astronomy, 15, 387, \dodoi{10.1007/BF02714823}

\bibitem[{Armstrong {et~al.}(1990)Armstrong, Coles, Rickett, \&
  Kojima}]{Armstrong1990}
Armstrong, J.~W., Coles, W.~A., Rickett, B.~J., \& Kojima, M. 1990, The
  Astrophysical Journal, 358, 685, \dodoi{10.1086/169022}

\bibitem[{Bain {et~al.}(2012)Bain, Krucker, Glesener, \& Lin}]{Bain2012}
Bain, H.~M., Krucker, S., Glesener, L., \& Lin, R.~P. 2012, Astrophysical
  Journal, 750, 44, \dodoi{10.1088/0004-637X/750/1/44}

\bibitem[{Bastian(1994)}]{Bastian1994}
Bastian, T.~S. 1994, The Astrophysical Journal, 426, 774,
  \dodoi{10.1086/174114}

\bibitem[{Baumbach(1937)}]{Baumbach1937}
Baumbach, S. 1937, Astronomische Nachrichten, 263, 121,
  \dodoi{10.1002/asna.19372630602}

\bibitem[{Berghmans {et~al.}(2006)Berghmans, Hochedez, Defise, Lecat, Nicula,
  Slemzin, Lawrence, Katsyiannis, van~der Linden, Zhukov, Clette, Rochus, Mazy,
  Thibert, Nicolosi, Pelizzo, Sch{\"{u}}hle, Berghmans, Hochedez, Defise,
  Lecat, Nicula, Slemzin, Lawrence, Katsyiannis, van~der Linden, Zhukov,
  Clette, Rochus, Mazy, Thibert, Nicolosi, Pelizzo, \&
  Sch{\"{u}}hle}]{Berghmans2006}
Berghmans, D., Hochedez, J.~F., Defise, J.~M., {et~al.} 2006, AdSpR, 38, 1807,
  \dodoi{10.1016/J.ASR.2005.03.070}

\bibitem[{Carley {et~al.}(2013)Carley, Long, Byrne, Zucca, Shaun~Bloomfield,
  McCauley, \& Gallagher}]{Carley2013}
Carley, E.~P., Long, D.~M., Byrne, J.~P., {et~al.} 2013, Nature Physics, 9,
  811, \dodoi{10.1038/nphys2767}

\bibitem[{Cho {et~al.}(2007)Cho, Gary, Lee, Moon, \& Park}]{Cho2007}
Cho, K.-S., Gary, D.~E., Lee, J., Moon, Y.-J., \& Park, Y.~D. 2007, The
  Astrophysical Journal, 665, 799, \dodoi{10.1086/519160}

\bibitem[{Chrysaphi {et~al.}(2018)Chrysaphi, Kontar, Holman, \&
  Temmer}]{Chrysaphi2018}
Chrysaphi, N., Kontar, E.~P., Holman, G.~D., \& Temmer, M. 2018, The
  Astrophysical Journal, 868, 10, \dodoi{10.3847/1538-4357/aae9e5}

\bibitem[{Chrysaphi {et~al.}(2020)Chrysaphi, Reid, \& Kontar}]{Chrysaphi2020}
Chrysaphi, N., Reid, H. A.~S., \& Kontar, E.~P. 2020, The Astrophysical
  Journal, 893, 115, \dodoi{10.3847/1538-4357/ab80c1}

\bibitem[{Dauphin {et~al.}(2006)Dauphin, Vilmer, \& Krucker}]{Daupin2006}
Dauphin, C., Vilmer, N., \& Krucker, S. 2006, Astronomy {\&} Astrophysics, 455,
  339, \dodoi{10.1051/0004-6361:20054535}

\bibitem[{Eselevich {et~al.}(2015)Eselevich, Eselevich, Sadykov, \&
  Zimovets}]{Eselevich2015}
Eselevich, V.~G., Eselevich, M.~V., Sadykov, V.~M., \& Zimovets, I.~V. 2015,
  Advances in Space Research, 56, 2793, \dodoi{10.1016/j.asr.2015.03.041}

\bibitem[{Fokker \& D.(1963)}]{Fokker1963}
Fokker, A., \& D., A. 1963, Space Science Reviews, 2, 70,
  \dodoi{10.1007/BF00174028}

\bibitem[{Gordovskyy {et~al.}(2019)Gordovskyy, Kontar, Browning, \&
  Kuznetsov}]{Gordovskyy2019}
Gordovskyy, M., Kontar, E., Browning, P., \& Kuznetsov, A. 2019, The
  Astrophysical Journal, 873, 48, \dodoi{10.3847/1538-4357/ab03d8}

\bibitem[{Grechnev {et~al.}(2018)Grechnev, Lesovoi, Kochanov, Uralov,
  Altyntsev, Gubin, Zhdanov, Ivanov, Smolkov, \& Kashapova}]{Grechnev2018}
Grechnev, V.~V., Lesovoi, S.~V., Kochanov, A.~A., {et~al.} 2018, Journal of
  Atmospheric and Solar-Terrestrial Physics, 174, 46,
  \dodoi{10.1016/j.jastp.2018.04.014}

\bibitem[{Harvey {et~al.}(1996)Harvey, Hill, Hubbard, Kennedy, Leibacher,
  Pintar, Gilman, Noyes, Title, Toomre, Ulrich, Bhatnagar, Kennewell,
  Marquette, Patr{\'{o}}n, Sa{\'{a}}, \& Yasukawa}]{Harvey1996}
Harvey, J.~W., Hill, F., Hubbard, R.~P., {et~al.} 1996, Science, 272, 1284,
  \dodoi{10.2307/2889785}

\bibitem[{Ingale {et~al.}(2015)Ingale, Subramanian, \& Cairns}]{Ingale2015}
Ingale, M., Subramanian, P., \& Cairns, I. 2015, Monthly Notices of the Royal
  Astronomical Society, 447, 3486, \dodoi{10.1093/mnras/stu2703}

\bibitem[{Kahler {et~al.}(2019)Kahler, Ling, \& Gopalswamy}]{Kahler2019}
Kahler, S.~W., Ling, A.~G., \& Gopalswamy, N. 2019, Solar Physics, 294, 1,
  \dodoi{10.1007/s11207-019-1518-3}

\bibitem[{Kai \& McLean(1968)}]{Kai1968}
Kai, K., \& McLean, D.~J. 1968, Publications of the Astronomical Society of
  Australia, 1, 141, \dodoi{10.1017/s1323358000011097}

\bibitem[{Klein {et~al.}(1999)Klein, Khan, Vilmer, Delouis, \&
  Aurass}]{Klein1999}
Klein, K.~L., Khan, J.~I., Vilmer, N., Delouis, J.~M., \& Aurass, H. 1999,
  Astronomy and Astrophysics, 346

\bibitem[{Kontar {et~al.}(2017)Kontar, Yu, Kuznetsov, Emslie, Alcock, Jeffrey,
  Melnik, Bian, \& Subramanian}]{Kontar2017}
Kontar, E.~P., Yu, S., Kuznetsov, A.~A., {et~al.} 2017, Nature Communications,
  8, 1, \dodoi{10.1038/s41467-017-01307-8}

\bibitem[{Kontar {et~al.}(2019)Kontar, Chen, Chrysaphi, Jeffrey, Emslie,
  Krupar, Maksimovic, Gordovskyy, \& Browning}]{Kontar2019}
Kontar, E.~P., Chen, X., Chrysaphi, N., {et~al.} 2019, The Astrophysical
  Journal, 884, 122, \dodoi{10.3847/1538-4357/ab40bb}

\bibitem[{Lemen(2012)}]{Lemen2012}
Lemen, J.~R. 2012, Solar Phys, 275, 17, \dodoi{10.1007/s11207-011-9776-8}

\bibitem[{Lengyel-Frey {et~al.}(1985)Lengyel-Frey, Stone, \&
  Bougeret}]{LengyelFrey1985}
Lengyel-Frey, D., Stone, R.~G., \& Bougeret, J.~L. 1985, A{\&}A, 151, 215.
\newblock \url{https://ui.adsabs.harvard.edu/abs/1985A&A...151..215L/abstract}

\bibitem[{Liepmann {et~al.}(1957)Liepmann, Roshko, Liepmann, \&
  Roshko}]{Liepmann1957}
Liepmann, H., Roshko, A., Liepmann, H., \& Roshko, A. 1957, {Elements of
  gasdynamics}.
\newblock \url{https://ui.adsabs.harvard.edu/abs/1957elga.book.....L/abstract}

\bibitem[{Maguire {et~al.}(2020)Maguire, Carley, McCauley, \&
  Gallagher}]{Maguire2020}
Maguire, C.~A., Carley, E.~P., McCauley, J., \& Gallagher, P.~T. 2020,
  Astronomy {\&} Astrophysics, 633, A56, \dodoi{10.1051/0004-6361/201936449}

\bibitem[{Mancuso {et~al.}(2019)Mancuso, Frassati, Bemporad, \&
  Barghini}]{Mancuso2019}
Mancuso, S., Frassati, F., Bemporad, A., \& Barghini, D. 2019, A{\&}A, 624, 2,
  \dodoi{10.1051/0004-6361/201935157}

\bibitem[{Mann {et~al.}(1999)Mann, Jansen, Macdowall, Kaiser, \&
  Stone}]{Mann1999b}
Mann, G., Jansen, F., Macdowall, R.~J., Kaiser, M.~L., \& Stone, R.~G. 1999, {A
  heliospheric density model and type III radio bursts}, Tech. rep.
\newblock
  \url{https://ui.adsabs.harvard.edu/abs/1999A%26A...348..614M/abstract}

\bibitem[{Maxwell {et~al.}(1985)Maxwell, Dryer, \& McIntosh}]{Maxwell1985}
Maxwell, A., Dryer, M., \& McIntosh, P. 1985, Solar Physics, 97, 401,
  \dodoi{10.1007/BF00165999}

\bibitem[{Melnik {et~al.}(2018)Melnik, Brazhenko, Frantsuzenko, Dorovskyy, \&
  Rucker}]{Melink2018}
Melnik, V.~N., Brazhenko, A.~I., Frantsuzenko, A.~V., Dorovskyy, V.~V., \&
  Rucker, H.~O. 2018, Solar Physics, 293, \dodoi{10.1007/s11207-017-1234-9}

\bibitem[{Melrose(1975)}]{Melrose1975}
Melrose, D.~B. 1975, Solar Physics, 43, 79, \dodoi{10.1007/BF00155144}

\bibitem[{Morosan {et~al.}(2018)Morosan, Gallagher, Zucca, O, Fallows,
  Magdaleni{\'{c}}, Mann, Bisi, Kerdraon, Morosan, Gallagher, Zucca, O,
  Fallows, Morosan, Gallagher, Zucca, Flannagain, Fallows, Reid, \&
  Magdaleni}]{Morosan2018}
Morosan, D.~E., Gallagher, P.~T., Zucca, P., {et~al.} 2018

\bibitem[{Nelson \& Robinson(1975)}]{NelsonRob1975}
Nelson, G., \& Robinson, R.~D. 1975, Publications of the Astronomical Society
  of Australia, 2, 370, \dodoi{10.1017/s1323358000014363}

\bibitem[{Nelson \& Melrose(1985)}]{Nelson1985a}
Nelson, G.~J., \& Melrose, D.~B. 1985, in Solar radiophysics: Studies of
  emission from the sun at metre wavelengths, 333--359.
\newblock \url{http://adsabs.harvard.edu/abs/1985srph.book..333N}

\bibitem[{Nelson \& Sheridan(1974)}]{NelsonSheridan1974}
Nelson, G.~J., \& Sheridan, K.~V. 1974, in Coronal Disturbances (Springer
  Netherlands), 345--348, \dodoi{10.1007/978-94-010-2257-6{\_}39}

\bibitem[{Nindos {et~al.}(2011)Nindos, Alissandrakis, Hillaris, \&
  Preka-Papadema}]{Nindos2011}
Nindos, A., Alissandrakis, C.~E., Hillaris, A., \& Preka-Papadema, P. 2011,
  Astronomy {\&} Astrophysics, Volume 531, id.A31, 12 pp., 531,
  \dodoi{10.1051/0004-6361/201116799}

\bibitem[{Pesnell(2015)}]{Pesnell2015}
Pesnell, W.~D. 2015, Handbook of Cosmic Hazards and Planetary Defense, 179,
  \dodoi{10.1007/978-3-319-03952-7{\_}16}

\bibitem[{Pomoell {et~al.}(2008)Pomoell, Vainio, \& Pohjolainen}]{Pomoell2008}
Pomoell, J., Vainio, R., \& Pohjolainen, S. 2008, Proceedings of the
  International Astronomical Union, 4, 493, \dodoi{10.1017/s1743921309029767}

\bibitem[{Riddle(1974)}]{Riddle1974}
Riddle, A.~C. 1974, Solar Physics, 35, 153, \dodoi{10.1007/BF00156964}

\bibitem[{Saito {et~al.}(1970)Saito, Makita, Nishi, \& Hata}]{Saito1970}
Saito, K., Makita, M., Nishi, K., \& Hata, S. 1970, AnTok, 12, 51.
\newblock \url{https://ui.adsabs.harvard.edu/abs/1970AnTok..12...53S/abstract}

\bibitem[{Schmidt {et~al.}(2016)Schmidt, Cairns, Gopalswamy, \&
  Yashiro}]{Schmidt2016}
Schmidt, J.~M., Cairns, I.~H., Gopalswamy, N., \& Yashiro, S. 2016, Journal of
  Geophysical Research: Space Physics, 121, 9299,
  \dodoi{10.1002/2016JA022956.Abstract}

\bibitem[{Sheridan {et~al.}(1972)Sheridan, Labrum, \& Payten}]{Sheridan1972}
Sheridan, K.~V., Labrum, N.~R., \& Payten, W.~J. 1972, Nature Physical Science,
  238, 115, \dodoi{10.1038/physci238115a0}

\bibitem[{Stansby {et~al.}(2020)Stansby, Yeates, \& Badman}]{Stansby2019}
Stansby, D., Yeates, A., \& Badman, S. 2020, Journal of Open Source Software,
  5, 2732, \dodoi{10.21105/joss.02732}

\bibitem[{Steinberg {et~al.}(1971)Steinberg, Aubier-Giraud, Leblanc, \&
  Boischot}]{Steinberg1971}
Steinberg, J.~L., Aubier-Giraud, M., Leblanc, Y., \& Boischot, A. 1971, A{\&}A,
  10, 362.
\newblock \url{https://ui.adsabs.harvard.edu/abs/1971A&A....10..362S/abstract}

\bibitem[{Stewart(1972)}]{Stewart1972}
Stewart, R.~T. 1972, Publications of the Astronomical Society of Australia, 2,
  100, \dodoi{10.1017/s1323358000013059}

\bibitem[{Suzuki {et~al.}(1985)Suzuki, Dulk, Suzuki, \& Dulk}]{Suzuki1985}
Suzuki, S., Dulk, G.~A., Suzuki, S., \& Dulk, G.~A. 1985, srph, 289.
\newblock \url{https://ui.adsabs.harvard.edu/abs/1985srph.book..289S/abstract}

\bibitem[{Thejappa \& MacDowall(2008)}]{Thejappa2008}
Thejappa, G., \& MacDowall, R.~J. 2008, The Astrophysical Journal, 676, 1338,
  \dodoi{10.1086/528835}

\bibitem[{Thejappa {et~al.}(2007)Thejappa, MacDowall, \& Kaiser}]{Thejappa2007}
Thejappa, G., MacDowall, R.~J., \& Kaiser, M.~L. 2007, The Astrophysical
  Journal, 671, 894, \dodoi{10.1086/522664}

\bibitem[{van Diepen {et~al.}(2018)van Diepen, Dijkema, \&
  Offringa}]{vanDiepen2018}
van Diepen, G., Dijkema, T.~J., \& Offringa, A. 2018, Astrophysics Source Code
  Library, 1804.003.
\newblock \url{https://ui.adsabs.harvard.edu/abs/2018ascl.soft04003V/abstract}

\bibitem[{van Haarlem {et~al.}(2013)van Haarlem, Wise, Gunst, Heald, McKean,
  Hessels, de~Bruyn, Nijboer, Swinbank, Fallows, Brentjens, Nelles, Beck,
  Falcke, Fender, H{\"{o}}randel, Koopmans, Mann, Miley, R{\"{o}}ttgering,
  Stappers, Wijers, Zaroubi, Akker, Alexov, Anderson, Anderson, van Ardenne,
  Arts, Asgekar, Avruch, Batejat, B{\"{a}}hren, Bell, Bell, van Bemmel,
  Bennema, Bentum, Bernardi, Best, B{\^{i}}rzan, Bonafede, Boonstra, Braun,
  Bregman, Breitling, van~de Brink, Broderick, Broekema, Brouw, Br{\"{u}}ggen,
  Butcher, van Cappellen, Ciardi, Coenen, Conway, Coolen, Corstanje, Damstra,
  Davies, Deller, Dettmar, van Diepen, Dijkstra, Donker, Doorduin, Dromer,
  Drost, van Duin, Eisl{\"{o}}ffel, van Enst, Ferrari, Frieswijk, Gankema,
  Garrett, de~Gasperin, Gerbers, de~Geus, Grie{\ss}meier, Grit, Gruppen,
  Hamaker, Hassall, Hoeft, Holties, Horneffer, van~der Horst, van Houwelingen,
  Huijgen, Iacobelli, Intema, Jackson, Jelic, de~Jong, Juette, Kant,
  Karastergiou, Koers, Kollen, Kondratiev, Kooistra, Koopman, Koster,
  Kuniyoshi, Kramer, Kuper, Lambropoulos, Law, van Leeuwen, Lemaitre, Loose,
  Maat, Macario, Markoff, Masters, McKay-Bukowski, Meijering, Meulman, Mevius,
  Middelberg, Millenaar, Miller-Jones, Mohan, Mol, Morawietz, Morganti,
  Mulcahy, Mulder, Munk, Nieuwenhuis, van Nieuwpoort, Noordam, Norden, Noutsos,
  Offringa, Olofsson, Omar, Orr{\'{u}}, Overeem, Paas, Pandey-Pommier, Pandey,
  Pizzo, Polatidis, Rafferty, Rawlings, Reich, de~Reijer, Reitsma, Renting,
  Riemers, Rol, Romein, Roosjen, Ruiter, Scaife, van~der Schaaf, Scheers,
  Schellart, Schoenmakers, Schoonderbeek, Serylak, Shulevski, Sluman, Smirnov,
  Sobey, Spreeuw, Steinmetz, Sterks, Stiepel, Stuurwold, Tagger, Tang, Tasse,
  Thomas, Thoudam, Toribio, van~der Tol, Usov, van Veelen, van~der Veen, ter
  Veen, Verbiest, Vermeulen, Vermaas, Vocks, Vogt, de~Vos, van~der Wal, van
  Weeren, Weggemans, Weltevrede, White, Wijnholds, Wilhelmsson, Wucknitz,
  Yatawatta, Zarka, Zensus, \& van Zwieten}]{vanHaarlem2013}
van Haarlem, M.~P., Wise, M.~W., Gunst, A.~W., {et~al.} 2013,
  \dodoi{10.1051/0004-6361/201220873}

\bibitem[{Vr{\v{s}}nak {et~al.}(2001)Vr{\v{s}}nak, Aurass, Magdaleni{\'{c}}, \&
  Gopalswamy}]{Vrsnak2001}
Vr{\v{s}}nak, B., Aurass, H., Magdaleni{\'{c}}, J., \& Gopalswamy, N. 2001,
  Astronomy {\&} Astrophysics, 377, 321, \dodoi{10.1051/0004-6361:20011067}

\bibitem[{Vr{\v{s}}nak \& Cliver(2008)}]{VrsnakCliver2008}
Vr{\v{s}}nak, B., \& Cliver, E.~W. 2008, Solar Physics, 253, 215,
  \dodoi{10.1007/s11207-008-9241-5}

\bibitem[{Vr{\v{s}}nak {et~al.}(2002)Vr{\v{s}}nak, Magdalenic, Aurass, \&
  Mann}]{Vrsnak_Mag2002}
Vr{\v{s}}nak, B., Magdalenic, J., Aurass, H., \& Mann, G. 2002, A{\&}A, 396,
  673, \dodoi{10.1051/0004-6361:20021413}

\bibitem[{Wang {et~al.}(2017)Wang, Liu, Wang, Hu, Shen, Jiang, \&
  Zhu}]{Wang2017}
Wang, W., Liu, R., Wang, Y., {et~al.} 2017, Nature Communications, 8, 1,
  \dodoi{10.1038/s41467-017-01207-x}

\bibitem[{Wuelser {et~al.}(2004)Wuelser, Lemen, Tarbell, Wolfson, Cannon,
  Carpenter, Duncan, Gradwohl, Meyer, Moore, Navarro, Pearson, Rossi, Springer,
  Howard, Moses, Newmark, Delaboudiniere, Artzner, Auchere, Bougnet, Bouyries,
  Bridou, Clotaire, Colas, Delmotte, Jerome, Lamare, Mercier, Mullot, Ravet,
  Song, Bothmer, \& Deutsch}]{Wuelser2004}
Wuelser, J.-P., Lemen, J.~R., Tarbell, T.~D., {et~al.} 2004, Telescopes and
  Instrumentation for Solar Astrophysics, 5171, 111, \dodoi{10.1117/12.506877}

\bibitem[{Zhang {et~al.}(2020)Zhang, Zucca, Sridhar, Wang, Bisi, Dabrowski,
  Krankowski, Mann, Magdalenic, Morosan, \& Vocks}]{Zhang2020}
Zhang, P., Zucca, P., Sridhar, S.~S., {et~al.} 2020, Astronomy {\&}
  Astrophysics, 639, A115, \dodoi{10.1051/0004-6361/202037733}

\bibitem[{Zimovets {et~al.}(2012)Zimovets, Vilmer, Chian, Sharykin, \&
  Struminsky}]{Zimovets2012}
Zimovets, I., Vilmer, N., Chian, A. C.~L., Sharykin, I., \& Struminsky, A.
  2012, Astronomy {\&} Astrophysics, Volume 547, id.A6 13 pp., 547,
  \dodoi{10.1051/0004-6361/201219454}

\bibitem[{Zucca {et~al.}(2014)Zucca, Carley, Bloomfield, \&
  Gallagher}]{Zucca2014}
Zucca, P., Carley, E.~P., Bloomfield, D.~S., \& Gallagher, P.~T. 2014,
  \dodoi{10.1051/0004-6361/201322650}

\bibitem[{Zucca {et~al.}(2018)Zucca, Morosan, Rouillard, Fallows, Gallagher,
  Magdalenic, Klein, Mann, Vocks, Carley, Bisi, Kontar, Rothkaehl, Dabrowski,
  Krankowski, Anderson, Asgekar, Bell, Bentum, Best, Blaauw, Breitling,
  Broderick, Brouw, Br{\"{u}}ggen, Butcher, Ciardi, Geus, Deller, Duscha,
  Eisl{\"{o}}ffel, Garrett, Grie{\ss}meier, Gunst, Heald, Hoeft,
  H{\"{o}}randel, Iacobelli, Juette, Karastergiou, Leeuwen, McKay-Bukowski,
  Mulder, Munk, Nelles, Orru, Paas, Pandey, Pekal, Pizzo, Polatidis, Reich,
  Rowlinson, Schwarz, Shulevski, Sluman, Smirnov, Sobey, Soida, Thoudam,
  Toribio, Vermeulen, van Weeren, Wucknitz, \& Zarka}]{Zucca2018b}
Zucca, P., Morosan, D.~E., Rouillard, A.~P., {et~al.} 2018, Astronomy {\&}
  Astrophysics, 615, A89, \dodoi{10.1051/0004-6361/201732308}

\end{thebibliography}
\end{document}